\documentclass[prl,amsmath,amssymb,floatfix,superscriptaddress,twocolumn,showpacs,footinbib]{revtex4}
\usepackage{graphicx}
\usepackage{dcolumn}
\usepackage{bm}
\usepackage{latexsym}
\usepackage{amscd} 
\usepackage[bookmarks, colorlinks=true, plainpages = false, citecolor = green, urlcolor = blue, filecolor = blue]{hyperref} 

\newcommand{\be}{\begin{equation}}
\newcommand{\ee}{\end{equation}}

\begin{document}

\title{A model of human population motion}
\author{Joseph D. Skufca}
\email{jskufca@clarkson.edu}
\affiliation{Department of Mathematics, Clarkson University, Potsdam, NY 13699-5815, USA}%
\author{Daniel ben-Avraham}
\email{benavraham@clarkson.edu}
\affiliation{Department of Physics, Clarkson University, Potsdam, NY 13699-5820, USA}%
\affiliation{Department of Mathematics, Clarkson University, Potsdam, NY 13699-5815, USA}

\date{\today}

\begin{abstract}
We introduce a basic model for human mobility that accounts for the different dynamics arising from 
individuals embarking on short trips (and returning to their home locations) and individuals relocating to a new
home.  The two modes of motion occur on widely separated time scales and affect different phenomena: for example, short trips, constituting the bulk of human motion, dominate the spread of diseases, while the spread of genes or family names is more strongly affected by relocations.  In many everyday life situations the two modes of motion need to be considered simultaneously.  This is illustrated by a simple SIR model for epidemic spreading where trips and relocations lead to dramatic differences even under equal volumes of travel.
\end{abstract} 


\maketitle

We introduce a basic model for human mobility that accounts for the different dynamics arising from 
individuals embarking on short trips (and returning to their home locations) and individuals relocating to a new
home.  The differences between the two modes of motion comes to light on contrasting two recent studies, one tracking
the geographical location of dollar bills~\cite{brockmann}, the other that of mobile cell phones~\cite{gonzalez}.  Trips introduce
two characteristic time scales; the time between trips, $\theta$, and the duration of each trip, $\tau$, and relocations introduces a third time scale, $T$, for the time between relocations.  In practice, $T\sim{\rm years}$, $\theta\sim{\rm months}$, and $\tau\sim{\rm days}$, so the three time scales are widely separated.  Traditionally, studies incorporating
human motion assume only a single mode, using a generic rate to account for all types of motion.

In what follows, we first introduce a social mobility model, capable of incorporating both modes of motion ---  trips and relocations --- and show how it can be cast as a Markovian process, thus simplifying its analysis.
We then argue that the distinction between the two modes 
is more that simple semantics or a minor increase in resolution, but may have significant consequences for dynamics supported by the motion, such as the spreading of an epidemic.  We illustrate though a specific example where an infectious disease, evolving according to the susceptible-infected-recovered (SIR) model, and spreading between two populations through trips and relocations, exhibits dramatic differences in the endemic levels of disease, depending on the mode of travel.


We now describe our {\it social mobility model} (SMM),  formulated as a Markov process.  Assume that the human population may be viewed as a number of homogeneous agents, $M$, moving independently among $N$ discrete geographic locations.  Because agents move independently, we need only describe the process for a single individual.  To support preferential return to home, we define states of the system as ordered pairs 
\be \label{eqstatedefn}
(i,j)=(\makebox{current location}, \makebox{home location}).
\ee
Denote the fraction of individuals at state $(i,j)$ by $\pi_{ij}(t)$ and require, accordingly,
\be 
\sum_{i,j} \pi_{ij}(t) =1.
\ee
Under this conceptual description, 
\be 
\pi_{i\star}(t)=\sum_j \pi_{ij}(t) {\rm\  and\ } \pi_{\star j}(t)=\sum_i \pi_{ij}(t)
\ee
are the fraction of people actually at location $i$, and the fraction of people whose home is $j$, respectively. 
Note that $\pi_{\star j}$ would be expected to match census data, $\sigma_j$, taken at some time close to  the start of the process (at time $t=0$).
We may in fact assume that the census is a fair representation of the equilibrium distribution, and
\[
\sigma_j \approx \pi_{\star j}(0) \approx \pi_{\star j}(t),
\] 
where the approximations are based on the idea that for a large population, the stochastic deviation from equilibrium should be small.   Implicit is also the assumption that $\pi_{ij}\ll\sigma_j$ for all $i\neq j$.

The SMM allows for two types of motion; {\em relocation} to a new home and short {\em trips}.
Relocations are relatively rare, and we denote the typical time between any two relocations (for a given agent) by $T$.
Trips are more frequent and are characterized by the time between trips, $\theta$, and the typical time $\tau$
spent in one location while traveling.  In practice, these three time scale are well separated, $T\gg\theta\gg\tau$, as $T\sim{\rm years}$, $\theta\sim{\rm months}$, and $\tau\sim{\rm days}$, though this is not required by our model.

An individual at his home location, $i$, [state $(i,i),$] may decide to relocate to $j$, [state $(i,j)$,] at rate $\omega_{ij}/T$.  The factor $\omega_{ij}$ expresses the likelihood to relocate from $i$ to $j$ (as opposed to some other location) and is normalized as $\sum_i\sum_j\sigma_i\omega_{ij}=1$, such that $1/T$ is approximately the total rate for relocations.
Rather than relocating, he may decide to take a trip to location $j$, $(i,i)\to(j,i)$, at rate $\nu_{ij}/\theta$, where the $\nu_{ij}$ 
indicate the preference to travel from $i$ to $j$ and are normalized as $\sum_i\sum_j\sigma_i\nu_{ij}=1$, such that the total rate for trips is approximately $1/\theta$.

A person away from home, $j$, at state $(i,j)$, would stay at $i$ an average length of time $\tau$, whereupon he might {\em return} home, to $(j,j)$, with probability $r$ (and rate $r/\tau$), or continue on his trip, to $(k,j)$, with probability $1-r$ (and
rate $(1-r)\nu_{ik}/\tau$).  Note that in this fashion $r$ controls the length of a trip: if $r=1$ all trips are merely to a destination and back (no layovers), but more generally the average number of legs in a trip is $(1+r)/r$.

Finally, a person might decide to relocate while away from home, $(i,j)\to(i,k)$, at rate $\omega_{jk}/T$, but since $\pi_{i,j}\ll\sigma_j$ this is an uncommon event and not much is lost by neglecting this possibility.  We list it as an optional move.
The various transitions allowed by our model, and their rates, are summarized in Table~\ref{tarates}.

The $\omega_{ij}$'s and $\nu_{ij}$'s are assumed as input, though we acknowledge that obtaining specific values is
an art in its own right and may involve a great deal of modeling and data mining.  When such information is not available, a simple choice, emulating
the {\em gravity model}~\cite{bharti}, could be $\nu_{ij}=Av_{ij}$, where
\be \label{eqvij}
v_{ij}=\frac{\sigma_j}{d_{ij}^\gamma}, \qquad A^{-1}=\sum_i\sum_j\sigma_i v_{ij}\,,
\ee
for $i\neq j$, and $\nu_{ii}=0$.
The idea here is that the attractiveness of a place, $j$, as a trip destination (from $i$), is proportional to its population, $\sigma_j$, but inversely proportional to the  travel distance from $i$ to $j$, $d_{ij}$.  The exponent $\gamma$ could be set equal to 2, acknowledging Kleinberg's condition for maximum navigability of the network through a decentralized algorithm~\cite{kleinberg}, or to $\gamma=1$, so as to maximize the information entropy and the ease of collecting information in the social net~\cite{hu}, but other choices might be equally reasonable.  At this naive level, the same assumptions might be made for relocations, i.e., $\omega_{ij}=\nu_{ij}$.

\renewcommand{\arraystretch}{2.2}
\begin{table}
\begin{ruledtabular}
\begin{tabular*}{3.2in}{c@{\extracolsep{\fill}}   c@{\extracolsep{\fill}}c}
{\bf transition} & {\bf rate} & {\bf description} \\ \hline
\multicolumn{3}{c}{\underline{When at home\!\!\!\phantom{y}}}\\
$(i,i)  \to  (j,i)$ & ${\displaystyle \frac{1}{\theta}\nu_{ij}}$ & travel \\
$(i,i)  \to  (i,j)$ & ${\displaystyle \frac{1}{T}\omega_{ij}}$ & relocate \\
\multicolumn{3}{c}{\underline{When away from home}}\\
$(i,j)  \to  (j,j)$ & ${\displaystyle \frac{r}{\tau}}$ & return home \\
$(i,j)  \to  (k,j)$ & ${\displaystyle \frac{1-r}{\tau}} \nu_{ik}$ & continue trip \\
$(i,j)  \to  (i,k)$ & ${\displaystyle \frac{1}{T}} \omega_{jk}$ & relocate (optional) \\ 
\end{tabular*}
\end{ruledtabular}
\caption{Transition rates for the SSM.  In all cases, $i\neq j\neq k$.}
\label{tarates}
\end{table}

The SMM can be simulated as a Markov process in a computer, using the rates of Table~\ref{tarates}, or it can be analyzed
through a master equation, incorporating the very same rates.  At the mean-field level, taking populations as  continuous variables and neglecting fluctuations in number space, the process is described by the rate equations:
\begin{subequations}
\label{eqrateeqn}
\begin{eqnarray} 
\dot{\pi}_{ii} &=&  \frac{r}{\tau} \sum_{j\neq i} \pi_{ji}-\left( \frac{1}{\theta}\sum_j \nu_{ij} + \frac{1}{T}\sum_j \omega_{ij} \right) \pi_{ii} \nonumber \\
& & \underbrace{+\frac{1}{T}\sum_j \omega_{ji}\pi_{ij}}_{\text{optional}}, \\
\dot{\pi}_{ij}&=&\frac{1}{\theta}\nu_{ji}\pi_{jj}+\frac{1}{T}\omega_{ij}\pi_{ii}+\frac{1-r}{\tau}\sum_{k \neq j} \nu_{ki}\pi_{kj} \nonumber \\
& &  -\left(\frac{r}{\tau} +\frac{1-r}{\tau}\sum_{k \neq j} \nu_{ik}\right) \pi_{ij}  \nonumber \\
& & + \underbrace{\frac{1}{T} \sum_k \left( \omega_{kj}\pi_{ik}-\omega_{jk}\pi_{ij} \right).}_{\text{optional}}  
\end{eqnarray}
\end{subequations}

We now turn to a particular implementation of the SMM that highlights the dramatic implications of its two modes of travel.
Consider a basic {\em susceptible-infected-recovered} (SIR) model with births and deaths occurring at equal rate, $\mu$, such that the total population, $N$, remains fixed (on average).  If all births are assumed to be susceptibles, then the governing equations are
\begin{equation}{\label{eqSdot}}
\begin{split}
&\dot{S}=\mu N -\mu S - \beta SI,\cr
&\dot{I}=\beta SI - \alpha I -\mu I,\cr
\end{split}
\end{equation}
along with the algebraic constraint that $S+I+R=N$. Here $\alpha$ describes the disease recovery rate and $\beta$ is a transmission rate parameter.  This model provides a reasonable representation for a disease such as the measles \cite{AndersonMay}.
Standard analysis of (\ref{eqSdot}) shows that if 
\[
R_0:=\frac{\beta N}{\mu+\alpha} >1\,,
\]
the disease is endemic and persists at some constant level, but
if $R_0<1$ the disease-free state is globally attracting.
Assuming a specific disease and social setting, we may view $\alpha,$ $\beta,$ and $\mu$ as fixed and observe that whether the disease is endemic or dies out is determined by the population size, $N$.  

Using this basic mean-field as a starting point, a number of researchers have shown that a reasonable interpretation for many diseases is that large population centers may operate at endemic levels, while outlying rural areas (where the disease would normally die out) have sustained levels or recurrent epidemics due to transport from the population centers \cite{bartlett1956}.  Within that general construct, we examine a simple situation where we model a two-site system: site $a$, a city, and site $b$, a village, where (for specificity) we take the city population, $N_a$, to be a factor of ten larger than the village population, $N_b.$   Choosing appropriate parameters, we may consider a disease such that if two populations remain separate, the disease is endemic in the city and dies out in the village.  But if we allow a small amount of social transport between locations, then the disease becomes endemic in both locations.  We use this simple setting to show the dramatically different effects of the two types of motion; {\it relocations} and {\it trips}.

 Define our state variables as $X_{ij},$ the number of people in health condition $X \in \{S,I,R\}$  who are at location $i$ and call $j$ their home, where  $i,j \in \{a,b\}.$  Thus, for example, $S_{aa}$ are the number of susceptible people who are at the city $a$ and $a$ is their home.  Applying a mean-field version of the SMM to \eqref{eqSdot}   yields
 \begin{eqnarray}\label{eqSaadot}
\dot{S}_{aa} &=& \mu N_a - \mu S_{aa} -\beta S_{aa} \left(I_{aa}+I_{ab}\right)\nonumber\\
&&-\omega_{ab} S_{aa} - \nu_{ab} S_{aa} + \omega_{ba} S_{bb} +\frac{1}{\tau} S_{ba},
\end{eqnarray}
which we describe and justify as follows:
\begin{itemize}
\item All births occur while the person is home (not on a trip).
\item A susceptible is equally likely to get the disease from any infected person who is in the same location.
\item {\it Relocations}  for all health conditions occur at the same rate $\omega_{ab}$  (or $\omega_{ba}$) and, following the gravity motion model, we further assume $ N_a \omega_{ab}=  N_b.\omega_{ba}.$
\item {\it Trips}, for all health conditions, occur at rate $\nu_{ab}$ ($\nu_{ba}$), again satisfying the gravity constraint $N_a \nu_{ab}= N_b\nu_{ba}.$
\item We assume that relocations occur only from home.
\item The last four terms of \eqref{eqSaadot} describe motion. For instance, the very last term refers to people returning from a trip (whose average duration is $\tau$).
\item We assume that the motions imply a change of location or home, but not a change of health condition.
\end{itemize}
The rate equations for the remainder of the state variables are straightforward and follow the same principles.
Rather than listing the full complement, we give just two further examples:
\begin{eqnarray}\label{eqIaadot}
\dot{I}_{aa} &=& \beta S_{aa} \left(I_{aa}+I_{ab}\right) - \alpha I_{aa}-\mu I_{aa}\nonumber\\
&&-\omega_{ab} I_{aa} - \nu_{ab} I_{aa} + \omega_{ba} I_{bb} +\frac{1}{\tau} I_{ba},
\end{eqnarray}
and
\begin{equation} \label{eqSbadot}
\dot{S}_{ba} =  - \mu S_{ba} -\beta S_{ba} \left(I_{ba}+I_{bb}\right)
+ \nu_{ab} S_{aa} -\frac{1}{\tau} S_{ba},
\end{equation}
where \eqref{eqIaadot} re-emphasizes that transport preserves the health status, while \eqref{eqSbadot} highlights the rule that no {\em relocations} take place while not at home (state $X_{i,j\neq i}$).

We now fix our parameter set to
\[
N_a=10,\, N_b=1,\,  \mu=0.05,\,\,  \alpha=48,\, \beta=40,\,  \tau=5/365,
\]
which, according to the mean-field description, and in the absence of motion,  supports an endemic level in the city $a$, but not in the village $b$.  Apart from the average staying time on a trip, $\tau$ (which is fixed at
$\tau=5$ days), the motion is governed by the yet unspecified rates $\omega_{ab}$ and $\nu_{ab}$.  These
can be  constrained so that the total level of transport, from trips and relocations combined, is fixed at some
small constant value.  The relative volume of trips versus relocations is then controlled by a single
parameter
\[
V:=\frac{T}{\theta},
\]
that we term the population {\em vacation index}: If motion includes only relocations then $V=0$, and $V$ grows ever larger as the fraction of trips increases.
Our general concept that $T>\theta$ would indicate that normally $V>1,$ though one can certainly imagine cultures where this might not be the case.

\begin{figure}
\includegraphics[width=0.48\textwidth]{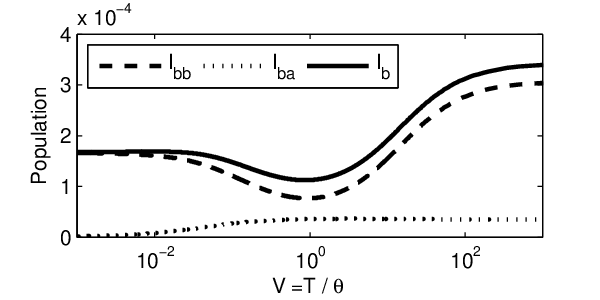}
\caption{Epidemic levels in the village as a function of the vacation index $V$.  Shown are $I_{bb}$ and
$I_{ba}$, as well as the total $I_{b}=I_{bb}+I_{ba}$.
\label{fig:figinfected}}
\end{figure}

Our SIR model has a single stable fixed point, which can be solved numerically.   
In Fig.~\ref{fig:figinfected} we show the steady-state one obtains in this fashion for the number of infected people in the village, as a function of the vacation index~$V$.  For ``normal" conditions, $V>1$, increased trips (offset by decreased relocations) results in a substantial increase of the disease level.  For $V<1$ (but increasing $V$), there is a decrease in the total number of infected individuals in the village. This results
from the fact that sick people, coming from the city to the village on a trip, are less effective at transmitting 
the disease than those traveling for relocation, due to their much shorter stay.  Trips become far more dangerous to the villagers when they don't compete with relocations ($V>1$).  

\begin{figure}
\includegraphics[width=0.48\textwidth]{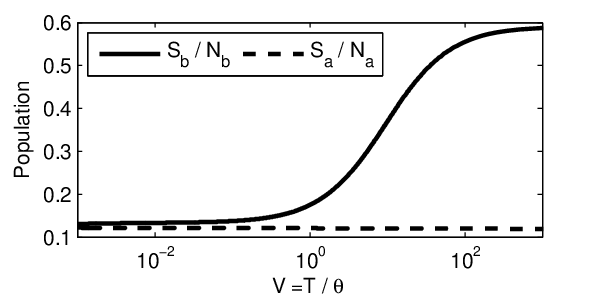}
\caption{Fraction of susceptibles, based at home.  Observe that for low vacation index, the populations are essentially mixed, but when transport is dominated by trips, the locations are strongly heterogeneous, with the village having significantly greater risk of experiencing an epidemic.
\label{fig:figsus}}
\end{figure}

In Fig.~\ref{fig:figsus} we show the total fraction of susceptibles in the city and in the village (relative to their total populations), under the same scenario.  In the city, the large population acts effectively as a buffer, 
resulting in a nearly constant fraction of susceptibles even as the vacation index changes from one extreme to another.   The fraction of susceptibles in the village is nearly equal to that in the city, for low vacation index.  However, as $V$ increases, there is a nearly four-fold increase in the fraction of susceptibles in the village, as trips become dominant.  

\begin{figure}
\includegraphics[width=0.48\textwidth]{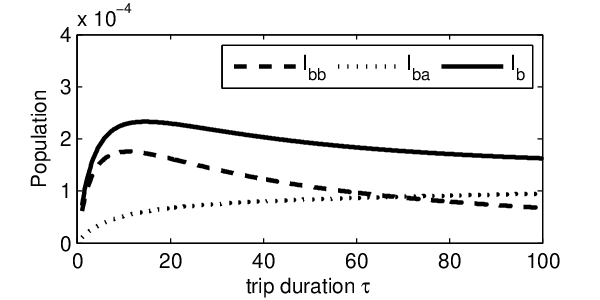}
\caption{Change in village infection population as a function of trip duration.  Fixing $V=10,$ we maintain transport volume while varying the trip duration, $\tau$.  The resonance peak in the endemic level at $\tau\approx 1/\alpha$ occurs only for a sizable fraction of trips (see text). 
\label{fig:figinftau}}
\end{figure}

So far, we have analyzed the effect of $T$ and $\theta$ alone, ignoring the important role of $\tau$, the third characteristic time scale of motion in the SMM.  In Fig.~\ref{fig:figinftau} we show the number of sick people in the village when the ratio of $T$ to $\theta$ is held fixed, at $V=10$, but the average duration of a trip, $\tau$, is varied from zero and up to 100 days.  One can see that peak transmission occurs when the trip duration
is roughly commensurate with the average length of the infection period, $\tau\approx1/\alpha=5\,$days.  Shorter trips
mean that infected individuals from the city return back home before fully spreading the disease, while longer trips allow villagers who travel to the city, and get infected there, to recover before returning home, thus reducing the level of disease in the village.  Notice that this interesting ``resonance" effect occurs only for trips,
and could not happen when relocations predominate (as confirmed by additional numerical analyses).

In summary, we have introduced a basic model for human mobility that incorporates two very distinct modes of motion;
trips and relocations, encompassing three widely different time-scales (two for trips and one for relocations).  All of these
time scales are potentially important, as different phenomena might evolve at  a characteristic time matching any of
the three.  Likewise, the fact that it is the very same individual that returns from a trip (as opposed to a randomly selected
one) has important ramifications for the spread of disease.  These hunches were demonstrated by a simple 
epidemic spreading SIR model evolving on just two locations, a ``city" and a ``village," showing that the endemic level
found in the village, of a disease such as the measles, depends strongly on the mode of travel (whether trips or relocations).  In addition, a resonance effect, where the endemic level maximizes, under unfavorable conditions, occurs
only when trips predominate.  Thus, making the distinction between the two modes of travel
leads to very different numerical results and reveals new qualitative phenomena.  For a fixed level of ``traffic'' between location, we observe that the trips provide an increase in the disease coupling (compared to the relocation-only pattern).  It is possible that this effect may contribute the required coupling parameter adjustment of a factor of 1.3 (as discussed in \cite{bharti}) to match England/Wales measles data to a gravity transport model.

Though couched in the specific language of human populations (and spread of disease, in our example), we anticipate that
similar considerations would  also apply elsewhere: animals, for example, might forage and
return to their home base (trips), or leave the pack to form a new clan (relocation), on widely different time scales, and this
might have implications to various aspects of their ecology, geographical genetic spread, etc.

Our model, while hopelessly naive, incorporates the crucial differences between the two modes of travel --- a distinction that to date has gone largely ignored.  Clearly, the model could be refined, by tweaking its rates, incorporating other modes of travel, and so on, and such future revisions will result in ever greater predictive power. 

\acknowledgments

We are grateful to the NSF, award PHY-0555312 (DbA), and DMS-0708083 (JDS), as well as NIH award R01 AG026553-01A2 (JDS), for partial support of this work.


\end{document}